# UTM Performance Under Stressing Scenarios

Ian Jessen
Transportation Safety and Resilience
MIT Lincoln Laboratory
Lexington, MA USA
ian.jessen@ll.mit.edu

*Abstract*—Proliferation of new classes of airspace participants, including uncrewed and advanced aerial mobility vehicles, necessitates the development and deployment of novel airspace management solutions, such as the Unmanned Traffic Management (UTM) system and the Provider of Services to UAM (PSU) Network. The efficacy of such systems has been demonstrated on multiple occasions via real-world deployments in limited test environments, however exploration of system behavior under stressing conditions requires the development of appropriate modeling and simulation (M&S) environments. Autonomy Networks for Advanced Mobility at Lincoln Laboratory (ANAMLL) is a virtual Systems Integration Laboratory (SIL) designed to host federated autonomy networks, such as a UTM or PSU Network, and to enable test and validation at scales not available in real-world deployments. As an example of ANAMLL's utility, we explore the performance of a representative UTM network during a stressing demand scenario. In a close examination of the demand scenario, ANAMLL demonstrates a UTM system demand point at which in-flight replanning can no longer be accomplished within an allowable time window. In a second analysis of the same scenario, ANAMLL demonstrates the impact of network connectivity performance on end-user airspace access.

## I. INTRODUCTION

Unmanned Traffic Management (UTM) systems are a critical enabler in the rapidly growing small Unmanned Aircraft Systems (sUAS) ecosystem. As drone technology becomes more widespread and accessible, the national airspace system—especially at low altitudes—faces increasing congestion and complexity. UTM is intended to address this challenge by providing the necessary architecture, protocols, and autonomy framework to manage the safe and efficient operation of multiple sUAS in shared airspace without human air traffic controllers.

As described in [1] UTM serves the following roles in facilitating safe, scalable, and integrated sUAS operations:

1. **Strategic Planning and Deconfliction:** Through pre-flight planning and real-time airspace updates, UTM helps prevent conflicts between participants by supporting intent discovery, sharing, and strategic deconfliction of planned operations.

2. **Situational Awareness:** UTM provides operators with real-time data about proximate operations and other supplemental airspace data, such as weather conditions.

3. **Communication and Notification:** UTM facilitates digital communication between service providers, operators, and regulators, enabling notifications about airspace changes and emergencies

4. **Compliance and Accountability:** UTM logs operations and helps ensure regulatory compliance, supporting post-event analysis and accountability in case of incidents.

UTM relies on a decentralized model, where multiple third-party UTM Service Suppliers (USSs) exchange information with each other and coordinate sUAS operations under a set of shared protocols and policies. UTM includes several infrastructural components to facilitate this information exchange, including a Discovery and Synchronization Service (DSS) and a Flight Information Management Service (FIMS). The DSS is responsible for maintaining a common airspace picture across all USS', while FIMS acts as gatekeeper and auditor of the UTM ecosystem while providing regulator access as well.

While UTM is targeted to the sUAS use case, a similar concept has been proposed for the management of Urban Aerial Mobility (UAM), involving larger and possibly passenger-carrying vehicles operating at low altitudes. This concept is referred to as a Provider of Services for UAM (PSU) Network, and broadly replicates the UTM network architecture as shown in [2]. Throughout this paper we focus specifically on the UTM use case, but the tools developed and resultant findings can be easily applied to the UAM use case as well

The development of UTM requires analysis of the safety, interoperability, and scalability of candidate solutions. Real-world testing is essential, but is limited by cost, time, and

DISTRIBUTION STATEMENT A. Approved for public release. Distribution is unlimited.

This material is based upon work supported by the Department of the Air Force under Air Force Contract No. FA8702-15-D-0001. Any opinions, findings, conclusions or recommendations expressed in this material are those of the author(s) and do not necessarily reflect the views of the Department of the Air Force.

safety considerations. As UTM systems evolve, modeling and simulation (M&S) will play a critical role in their development, validation, and certification. M&S tools enable researchers, developers, and regulators to evaluate UTM, derive or validate system performance requirements, and explore end-to-end performance under a range of scenarios without the cost and risk of live testing. Examples of such analyses include:

- Safety and efficiency under various UTM scenarios, system failure modes, and edge cases (e.g., communication loss, system latency, targeted cyberattacks).
- Performance of various conflict detection and resolution algorithms.
- Interoperability of UTM service providers (USS) in a federated environment.

Several UTM-focused M&S tools have been developed which focus on various aspects of end-to-end UTM efficacy. In [4] Zhao et al. present a fast-time simulation tool to study airspace traffic management policies and the relationship between low altitude demand/capacity and ground communication infrastructure demand/capacity. In [4], Hsieh et al. develop a simulation to assess protocols related to reasoning about 4-dimensional *Operational Volumes*. Still other efforts, such as those presented in [6] and [7] develop M&S environments that focus on the performance of specific routing or deconfliction algorithms. One aspect consistently left uncovered by M&S tools thus far is the role of the underlying communications architecture and protocols of any UTM implementation, especially at future levels of scale.

Scalability is one of the most crucial aspects of UTM—how the system performs as the number of service providers, users and operations increases. Simulation of UTM systems at scale is essential for stress testing the UTM architecture under hypothetical high-load conditions and evaluating the distributed algorithms for dynamic airspace allocation, conflict detection and resolution, and prioritization. To that end, MIT Lincoln Laboratory has developed a UTM simulation environment - Autonomy Networks for Advanced Mobility at Lincoln Laboratory (ANAMLL).

## II. METHODOLOGY

### A. Simulation Environment

The ANAMLL simulation environment provides a virtual system integration laboratory (SIL) for the assessment of UTM and similar federated autonomy networks. ANAMLL provides instrumented, parameterized, and extensible implementations of all components required to assemble a fully functional UTM "network-under-test", including USS, FIMS, and DSS. Component implementations are compliant with the standards of [5], typically written in Python 3.10 and adhere to industry best-practices for the development of high-performance web-based applications, such as microservice event-driven architectures and extensive use of cooperative concurrency. Where possible, algorithm implementation (e.g. strategic deconfliction strategies) are decoupled from underlying services, allowing desired algorithms to be specified as per-simulation parameters.

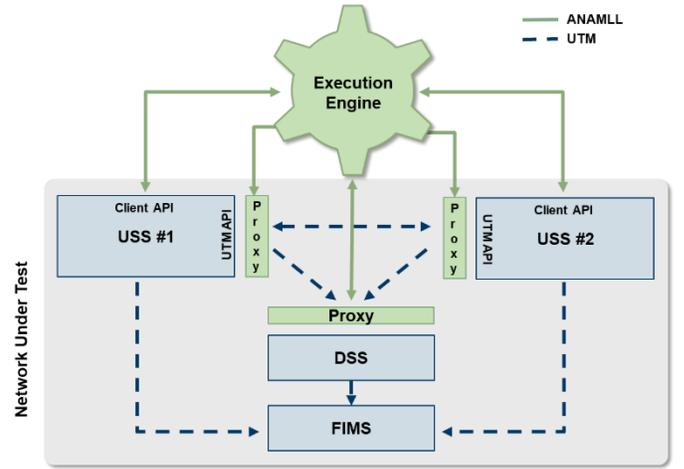

*Fig. 1 ANAMLL architecture with example network-under-test*

ANAMLL provides infrastructural components necessary to execute simulations and collect relevant data. Foremost among these is an "Execution Engine" which provides simulated inputs to the UTM network-under-test. During execution of a simulation the Execution Engine provides scripted inputs to the external-facing elements of a network-under-test, representative of operational and management interactions with UTM. These interactions include operation planning (strategic deconfliction and intent publication), flight execution (operational state changes and telemetry sharing), and airspace management (constraint publication and historical queries). The Execution Engine is also primarily responsible for collecting performance metrics during the course of a simulation, typically derived from the responses provided by the external-facing UTM interfaces of the network-under-test. Notably, the Execution Engine is responsive to the state of the network-under-test during simulation execution. An example simulation scenario may specify that an operation is submitted to UTM for deconfliction and sharing at $t_{sim}=60s$ and later activated for flight at $t_{sim}=360s$, however if the state of the network-under-test evolved such that the initial strategic deconfliction step failed due to an unresolvable conflict, the execution engine would not execute the later flight activation interaction.

The Execution Engine interfaces with interposing proxies that exist between specified pairs of connected components in a network-under-test. These proxies intercept and record all communications between connected components. Additionally, the proxies modify network connectivity properties over the course of a simulation. For example, a simulation scenario may dictate that at $t_{sim}=240s$ all connections inbound to a specific USS become highly unreliable (e.g. due to a simulated malicious distributed denial of service attack). To effect this, the Execution Engine sends a configuration message to the relevant proxy during simulation execution changing the appropriate model parameter. In this example, the interposing proxy would then fail incoming TCP connections with the specified random likelihood.

### B. Simulation Scenario

As an initial use case for exploring UTM system performance under stressing scenarios we have developed a simulation scenario derived from a hypothetical sequence of events occurring in a mid-term evolution of domestic low-altitude airspace. While this remains a hypothetical scenario, it was crafted with intentionally realistic assumptions about future development and should not strain credulity. Future scenario derivations will benefit from recent integrations with pedigreed demand models, such as [8].

The scenario captures a metro-area airspace managed by two separate USS providers. The representative network-under-test is shown in Fig. 1. At the beginning of the scenario, 40 pre-planned future operations are distributed across the two providers. As the scenario starts, a single airspace constraint is published that affects all the existing, pre-planned operations. This could be the result of a public-safety entity responding to an upcoming VIP visit, for example. As a result of the published constraint, all 40 pre-planned operations must be replanned. The USS's therefore experience a time-correlated spike in operational-planning tempo, including the strategic conflict detection (SCD) process.

Simultaneous with this planning-tempo spike, an inflight operation experiences an unanticipated condition which affects its ability to maintain conformance with its previously published operational intent (e.g., worse than expected winds aloft and/or delayed departure). The operator of the non-conformant operation must immediately replan and re-execute UTM-aided SCD. If conformity cannot be reestablished within 60s, the flight must declare a contingency and may have to be terminated early.

### C. Experimental Setup

Our simulation execution swept over two independent variables in the scenario described above. The first sweep adjusted the period during which the 40 operational replannings were submitted to UTM for SCD processing. As this window shrinks the spike in operational-planning becomes more severe and the instantaneous load on UTM grows, representative of higher levels of scale (e.g., more overall UTM users and/or greater levels of autonomy). The second swept variable adjusted the difference in latency between each USS and the single DSS node in the example network. Latency is a property of the network conditions experienced by each USS. Latency may grow as network conditions worsen (e.g., due to higher network load, possibly due to cyberattack), or latency can be minimized by competitive actions such as infrastructure relocation or other network topology manipulation.

Strategic deconfliction in the simulation scenario was implemented as a simple 4D conflict identification, with no attempt at either negotiation or rerouting. Operations were accepted only when all existing operations specified by the DSS were examined with no 4D conflicts identified. If the DSS common airspace picture changed between the start and end of any conflict detection, the process would be restarted using the newly updated airspace picture. This cycle would be attempted a maximum of 5 times per operation.

Simulations were executed on a dedicated host. Each ANAMLL component was deployed as a separate containerized application on the simulation host. Containers were interconnected via a dedicated virtualized network. The simulation host was equipped with dual Intel(R) Xeon(R) Gold 6254 CPUs, providing 36 physical cores operating at a base frequency of 3.10 GHz.

## III. RESULTS

Metrics were derived from Execution Engine interaction with the network-under-test during simulation. Of interest in the described scenario are two performance parameters associated with the two swept independent variables:

1. **SCD Process Duration:** time elapsed between submission of an operational intent by a USS user (i.e. Execution Engine) and subsequent determination by the USS whether the operation can be accepted into. The SCD process typically involves several spatial queries to geographic databases, network interactions between USS, DSS and possibly other USS', and 4D geometry analysis. The MIT LL USS provides push notifications of status updates, eliminating any influence of state polling. Values shown in Fig. 2 are the durations of the 95th percentile of all SCD operations per-USS over the course of repeated simulations of a constant parameter set.

2. **Operation Acceptance Rate:** frequency with which a given operation is accepted by its controlling USS (i.e. passes SCD without rejection) across repeated simulations of a constant parameter set. Fig. 3 shows the acceptance rate of a single operation on the Y-axis at each latency-degradation value where all other scenario parameters are kept constant.

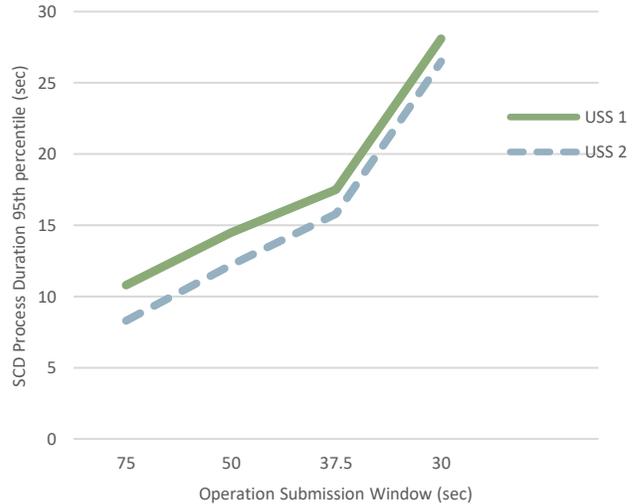

*Fig. 2 UTM Scalability*

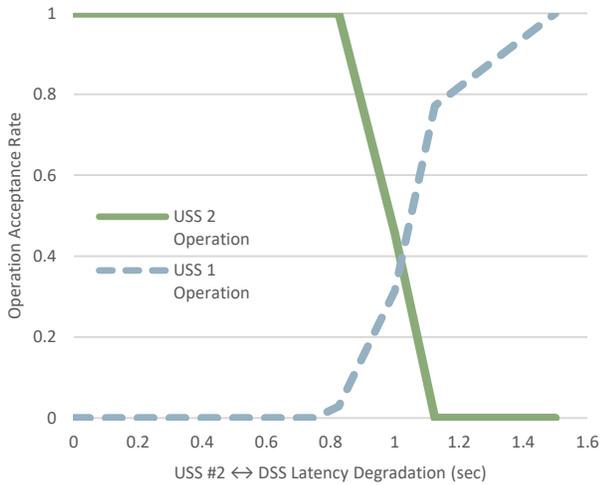

*Fig. 3 Equitable Airspace Access*

ANAMLL provides a real-time stochastic simulation environment. Individual executions of a constant scenario will vary due to uncontrolled factors within the execution environment. Pre-emptive process scheduling, instantaneous network conditions, and other similar external factors can and influence the results of each simulation execution. Results are therefore aggregated across numerous executions of each parameter set.

## IV. DISCUSSION

Our experimental design effectively addresses two questions related to overall UTM performance:

1. How does the existing UTM architectural design impact operational success at high levels of scale?
2. How does the existing UTM architectural design affect equitable airspace access?

In the first case we address scalability concerns with the existing UTM design. UTM relies on the Discovery and Synchronization Service to guarantee a common airspace picture among disparate USS providers. The DSS relies on "opaque version numbers" (OVN's) to ensure that each USS has utilized a provably current view of the relevant airspace in conducting SCD for a new or replanned operation. This approach enables *post facto* process synchronization across multiple USS's, ensuring that each accepted operation is provably deconflicted. This approach is susceptible to race conditions between USS's as they compete to submit proximate operations in a dynamic airspace. The sequence diagram of Fig. 4 illuminates such a race condition, where USS 2 completes conflict identification ahead of USS 1, despite Operator 1 having submitted their operations before Operator 2. As a result, the airspace picture available to USS 1 becomes outdated, forcing USS 1 to re-attempt the deconfliction process with a newly updated airspace picture.

The impact of the contentious submission strategy is frequent failed submissions and repetitive SCD processes during periods of high planning tempo. The repetition of SCD leads to increasing overall SCD process duration. Likelihood of race conditions increases in proportion to planning tempo, therefore correlating overall SCD process duration with planning tempo. In our simulation scenario SCD process durations approached 30 seconds in the moderate demand scenario of 40 operations submitted during a 30 second window.

The impact of this bottleneck could be significant for the inflight operation. Recall that the inflight operation in our simulation scenario must complete its replanning prior to the expiration of a 60 second grace period between losing operation intent conformance and declaring an operational contingency. In this example, nearly half that period is consumed solely by strategic conflict determination (driven by the race conditions described above), exclusive of actual route generation and optimization, operator acceptance, uplink, etc.

This result suggests that the contentious approach to airspace synchronization may not be appropriate for UTM SCD as scale increases. Future work with ANAMLL will investigate the introduction of basic synchronization primitives to the UTM protocol designed to address the needs of priority operational replanning.

The second question addresses the impact of current UTM design on equitable allocation of airspace among operators. This is a question intentionally separated from any DSS algorithms or other approaches designed to ensure equitable access, assuming only that the DSS provides first-come first-served operation acceptance as currently implemented.

That first-come first-served approach favors users who can communicate their operations to the DSS faster than others. This is particularly evident in our example scenario, when all pre-planned operations are replanned nearly simultaneously.

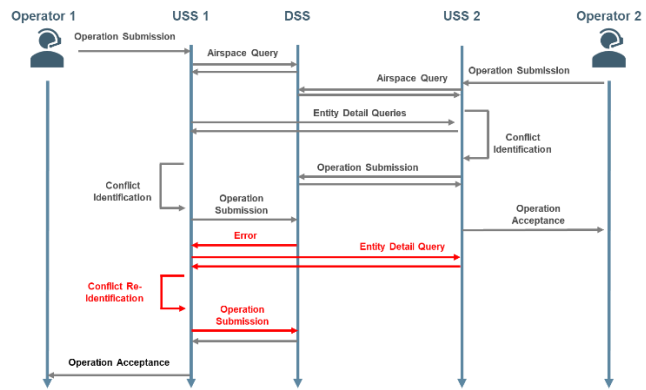

*Fig. 4 SCD sequence diagram with race condition*

In the sequence diagram of Fig. 4 we see the contention between two of these replanned operations, one serviced by each USS. The sequence diagram depicts 5 actors (shown as vertical lines) and the interactions between them (horizontal arrows) as time runs from the top of the diagram to the bottom. Notably, the operations submitted are not themselves in conflict, but rather have conflicts at one and even two levels of removal. At baseline, the operation serviced by USS 1 is always rejected. However, as the latency between USS 2 and the DSS is steadily increased, the likelihood of the USS 1 operation being accepted during the period of high planning-tempo increases. This demonstrates the susceptibility of the first-come first-served approach to poor network conditions, such as may be encountered during DDOS cyberattack. Alternatively, this demonstrates a competitive advantage to be gained by minimizing latency between a USS and the DSS.

## V. CONCLUSION

Work conducted to date demonstrates the importance of M&S tools for the future development of UTM and related airspace management. In this paper we have presented the ANAMLL simulation environment as a tool to assess performance of UTM as constrained by network architectures and protocols. Initial examinations of performance under a single stressing scenario suggests that a synchronization and/or prioritization scheme is necessary to avoid negative impacts of time-correlated demand spikes on the safety and efficiency of inflight operations. Additionally experimental results demonstrate the sensitivity of the UTM network architecture to competitive advantages or disadvantages in network connectivity and the impact on end-user airspace access.

The ANAMLL simulation environment has continued to evolve from the baseline presented above. Additional capabilities now include:

- Integration of offboard DAA solutions, such as described in [9]
- Simulation scenario generation leveraging formal demand models [8]
- Hardware-in-the-loop capability with multiple sUAS vehicles

Future work leveraging these new ANAMLL capabilities will focus on the validation of the simulation results as well as extending simulated scenarios to include the behavior of newly integrated services, such as DAA. This work will support follow on analyses of UTM and similar airspace management networks in support of defining and validating performance requirements for both regulator and implementers as these capabilities mature and provide support for airspace operations.